\documentclass[aps,prl,twocolumn,showpacs,floatfix,preprintnumbers,nofootinbib,superscriptaddress]{revtex4-1}

\usepackage{graphicx}
\usepackage{color}
\usepackage{natbib}
\usepackage{multirow}

\usepackage{epsfig}
\usepackage{graphicx}
\usepackage{amsfonts}
\usepackage{amssymb}
\usepackage{latexsym}
\usepackage{amsmath}
\usepackage{hyperref}

\usepackage{amsthm,amsmath,xcolor,latexsym,mathtools,hyperref,mathrsfs}
\usepackage{amssymb}
\usepackage{slashed}  
\usepackage{bm}
\usepackage{epsfig}

\newcommand{\be}{\begin{equation}}
\newcommand{\ee}{\end{equation}}
\newcommand{\beq}{\begin{equation}}
\newcommand{\eeq}{\end{equation}}
\newcommand{\ba}{\begin{eqnarray}}
\newcommand{\ea}{\end{eqnarray}}
\newcommand{\bef}{\begin{figure}}
\newcommand{\eef}{\end{figure}}

\newcommand{\p}{\partial}

\newcommand{\g}{\gamma}

\newcommand{\cL}{{\cal L}}

\begin{document}

\title{Soft gravitons as Goldstone modes of spontaneously broken asymptotic symmetries\\ in de Sitter spacetimes}
\author{Martin S. Sloth}
\email{sloth@sdu.dk}
\affiliation{Universe-Origins, University of Southern Denmark,\\ Campusvej 55, 5230 Odense M, Denmark}

\begin{abstract}
I demonstrate that soft graviton modes in de Sitter spacetimes are the Goldstone modes of the spontaneously broken asymptotic symmetry group of de Sitter space. I then show that any local measurement, including the effects of the environment, will collapse the symmetric state onto the broken state in the large volume limit. In any discussion involving observers, de Sitter spacetimes are, therefore, best described globally by the broken phase, while local observers, in the small volume limit, can not discriminate between different degenerate global vacuum states and are therefore best described by the symmetric state. As a consequence, a small Hubble-sized local region initially in the symmetric state will, after a time scale corresponding to the Page time of de Sitter space, have expanded to a large region in the broken state. This illuminates the physical nature of soft graviton modes in de Sitter spacetimes.
\end{abstract}

\maketitle

\section{Introduction}
There has been a long history of discussions of the physical reality of infrared super-horizon (soft) graviton modes in de Sitter spacetimes among practitioners of cosmology. The opinion ranges from arguments that they are pure gauge \cite{Allen:1986ta,Allen:1986tt,Higuchi:2000ye,Higuchi:2011vw}, with no physical meaning, to arguments that they are responsible for the ultimate demise of de Sitter space by the decay of the cosmological constant \cite{Polyakov:1982ug,Polyakov:2009nq,Polyakov:2007mm,Tsamis:1992sx,Tsamis:1994ca,Tsamis:1993ub,Tsamis:2007is,Myhrvold:1983hu,Dvali:2013eja,Dvali:2014gua,Dvali:2018jhn}
. In between, we find the more moderate point of view that they are a pure gauge locally but do have global effects once the gauge invariant observers are properly defined and included as part of the quantum system. The effects can be observed in the cosmic microwave background or large scale structure of the universe today when modes enter the horizon after the end of the quasi–de Sitter inflationary phase in the early universe or as deformations of the global properties of de Sitter spacetimes \cite{Giddings:2011zd,Giddings:2011ze,Giddings:2010nc,Ferreira:2016hee,Ferreira:2017ogo,Ferreira:2017erz}. Here I am going to further clarify the physical understanding of soft graviton modes by showing that they can be viewed as the Goldstone modes of spontaneously broken asymptotic symmetries in de Sitter space.

It was argued some years ago in the context of black holes that the Goldstone modes of the spontaneously broken asymptotic symmetries of the black holes, the Bondi-Metzner-Sachs (BMS) supertranslations, are the soft graviton modes satisfying Weinberg's soft graviton theorems \cite{He:2014laa, Hawking:2016msc}. There is a very analogous story in de Sitter space. The equivalent of the Weinberg soft theorem satisfied by the Weinberg adiabatic soft graviton modes is the Maldacena consistency relations, and the asymptotic symmetries of de Sitter space are the diffeomorphisms in three-dimensional Euclidean space \cite{Anninos:2010zf}. Here, I will show that the soft graviton modes are, indeed, also the Goldstone modes of the spontaneously broken asymptotic symmetry group of de Sitter space. In cosmology, we also think of soft graviton modes in de Sitter spacetimes as frozen-out super-horizon gravitational fluctuations, described by the traceless and transverse spatial component of the graviton. 

In order to show that the superhorizon soft graviton is the Goldstone of the spontaneously broken asymptotic symmetry group of de Sitter space, I will discuss the symmetry-breaking pattern and then show that the counting of Goldstone modes adds up correctly. At the linearized level, the anisotropic scaling symmetry generated by symmetric traceless matrices is broken, leaving five broken generators. However, in the case of broken space-time symmetries, it is well known that Goldstone's theorem for internal symmetries does not trivially apply. As in the case with spontaneously broken non-Lorentz invariant symmetries \cite{Nielsen:1975hm}, also in the case of space-time symmetries, the number of massless Goldstone modes can be different from the number of broken symmetry generators. In the case of unbroken translations, local translation in the unbroken directions can be used to relate the Goldstone modes to each other \cite{Ivanov:1975zq, Low:2001bw}. With three unbroken translations, only two physical Goldstone modes remain out of the five broken generators. As I will show below, these are the two massless Goldstone modes of the soft graviton, which remain as physical modes. This argument generalizes to any dimension.

In the next section, I discuss the symmetry-breaking pattern and how to count the Goldstone modes correctly. Then I discuss the role of local observers versus the role of global observers. Afterwards, I briefly discuss the generalization of these arguments to a period of slow-roll inflation where the time-translation invariance of de Sitter space is softly broken. Then, I discuss some examples from the literature and how they relate to the interpretation of soft gravitons as the Goldstone modes of spontaneously broken asymptotic symmetries. Finally, I conclude.

\section{Spontaneously broken asymptotic symmetries and soft modes as Goldstones}\label{spontaneously}
The solutions of Einstein's equations with a cosmological constant break the full diffeomorphism symmetry down to the isometry group of de Sitter space, $SO(D,1)$ in $D=1+n$ spacetime dimensions. However, an observer living on the space-like infinity of de Sitter space would experience a higher symmetry and conclude that the $SO(D,1)$ de Sitter invariance is enhanced to the asymptotic symmetry group of de Sitter space, which is the diffeomorphisms of ${\mathbb{R}}^n$ \cite{Anninos:2010zf}.

In $n$ Euclidian dimensions the general covariance implies that the index, $j$, up or down transforms as covariant or contravariant vectors under the linear matrix multiplication of matrices given by $\p x^{i}/\p  \tilde x^{j}$  $\p \tilde x^{j}/\p x^{i}$ at the level of linearized infinitesimal coordinate transformation. The linearized coordinate transformations are elements of the symmetry group $GL(n,\mathbb{R})$ with dimension $n^2$. $GL(n,\mathbb{R})$ can be written as $SL(n,\mathbb{R})\times\mathbb{R}$, where $SL(n,\mathbb{R})$ has $SO(n)$ as a subgroup with dimension $n(n-1)/2$ generated by the antisymmetric tensors $J_{ij}$, while the other $n(n+1)/2-1$ symmetric traceless generators $S_{ij}$ generate shear deformations. The dilatation, $D$, is part of $GL(n,\mathbb{R})$, but not $SL(n,\mathbb{R})$, which has dimension $n^2-1$. Including translation invariance enhances  $GL(n,\mathbb{R})$ to the affine group $Aff(n,R)) =  GL(n,\mathbb{R})\times P_n$, where $P_n$ is the group of translations. The affine group, therefore, has the Poincaré group as a subgroup. The full diffeomorphism group is the closure of the affine group and the special conformal group. Any generator of the full diffeomorphism group can be written in terms of (infinite) combinations of the generators of  $SL(n,\mathbb{R})$, as well as the generators of translations $P_n$, dilatations $D$ and special conformal transformations $K_n$ \cite{Borisov:1974bn}. Important for the discussion below, at the linearized level, the generators of the asymptotic symmetry group are therefore $S_{ij}, J_{ij}, P_i, K_i, D$.

When we restrict ourselves to pure de Sitter space (with unbroken time translation), the relevant gravitational degrees of freedom are contained in the transverse and traceless graviton $\gamma_{ij}$, which can be extracted from the metric in the planar slicing
\begin{eqnarray} 
ds^2=  -dt^2 +  a^2(t) \left[e^{\g}\right]_{ij} dx^i dx^j\, .
\end{eqnarray}
We have ignored the lapse and shift, which decay and, therefore, are unimportant on the asymptotic boundary.   The asymptotic symmetry transformations act on this metric as \cite{1203.6351}
\beq \label{metricvariation}
\delta \left(\left[e^{\gamma}\right]_{ij} \right)=\cL_\xi \left( \left[e^{\gamma}\right]_{ij}\right)~.
\eeq
Here $\cL_\xi$ is the Lie derivative with respect to the arbitrary spatial vector $\xi_i$.  

The variation of $\gamma$ under this transformation is to leading order
\beq \label{vgamma}
\delta\gamma_{ij}=\partial_i\xi_j+\partial_j\xi_i-\frac23\partial\cdot\xi\delta_{ij}+\mathcal{O}(\gamma).
\eeq
Clearly, diffeomorphisms that satisfy 
\begin{eqnarray}
\partial_i\xi_j+\partial_j\xi_i-\frac23\partial\cdot\xi\delta_{ij}=0
\label{x1}
\end{eqnarray}
can not generate tensor adiabatic modes. Equation (\ref{x1}) is solved by
\begin{eqnarray}
\xi_i=c_i+\omega_{ij}x_j+c x_i+\Big(x^2 a_i-2 x_i (\vec{x}\cdot \vec{a})\Big) \label{x2}
\end{eqnarray}
where $c_i$, $a_i$, $c$, and $\omega_{ij}=-\omega_{ji}$ are constant parameters. 
One recognizes (\ref{x2}) as the conformal Killing vector 
of $\mathbb{R}^3$, where $c_i$ is associated to the translations, $c$ to dilations, $a_i$ to conformal boosts and $\omega_{ij}$ to rotations. 

Therefore out of $SL(3)$, only its symmetric generators
\begin{eqnarray}
S_{ij}=x_i\partial_j+x_j \partial_i-\frac{2}{3} \delta_{ij}x^k\partial_k
\end{eqnarray} 
can be employed to generate a constant adiabatic mode 
\begin{eqnarray}
\delta \gamma_{ij}=s_{ij} \label{s1}
\end{eqnarray}
by 
\begin{eqnarray}
\xi_i=s_{ij}x_j
\end{eqnarray}
where $s_{ij}=s_{ji}$.

Let us consider a frozen-in soft graviton mode. It is the zero momentum limit of a plane wave model with momentum $k^L \to 0$. We can think of this mode as being a symmetric $3\times 3$ dimensional traceless matrix $\gamma_{ij}^L$ . This mode is generated by a linear diffeomorphism of the form
\beq\label{xiL}
\xi_i^L = -\frac{1}{2}\gamma_{ij}^L x^j
\eeq
implying $\delta \gamma_{ij} = -\gamma_{ij}^L$. This amounts to switching on a nonzero vacuum expectation value for the graviton $ \gamma^L_{ij} =\left<\gamma_{ij}(x)\right> $.

Therefore, in four space-time dimensions with $n=3$, we have that at the linearized level, the soft graviton mode is breaking $SL(3)$ down to $SO(3)$ and so one would expect a ${\rm dim}\, SL(3)/SO(3)=5$ Goldstone modes, corresponding to the five broken generators, $S_{ij}$, which is inconsistent with the number of physical graviton modes. The reason is that one may use the diffeomorphism group to cancel some of the five degrees of freedom of the adiabatic mode. It was shown in \cite{Ivanov:1975zq,Low:2001bw} that in the directions of unbroken spatial translations, one may use a local translation to remove the redundant degrees of freedom. This is only true when the commutator of the generators of the broken symmetries does not commute with the unbroken translation generators. That is why for internal symmetries where the generators of broken internal symmetries commute with the translations, this does not affect the counting of Goldstone modes, while in the case of broken space-time symmetries, which do not commute with the unbroken translation generators, the number of Goldstone modes is less than the number of broken generators.  Indeed, an $ x$-dependent 
translation is generated by 
\begin{eqnarray}
\xi_i=c_i(x) 
\end{eqnarray}
and generates the mode
\begin{eqnarray}
\delta \gamma_{ij}=\partial_i c_j(x)+\partial_j c_i(x)-
\frac{2}{3}\delta_{ij}\partial_kc_k(x). 
\end{eqnarray}
Clearly now, we can cancel  part of (\ref{s1}) by choosing 
\begin{eqnarray}
\partial_i c_j(x)+\partial_j c_i(x)-
\frac{2}{3}\delta_{ij}\partial_kc_k(x)=-s_{ij}
\end{eqnarray}
This is an equation that specifies the three functions $c_i(x)$, and therefore, we have only $5-3=2$ real Goldstone modes.  This is an example of the fact that when spacetime symmetries are involved, the usual rule that the number of Goldstones of the breaking $G\supset H$ is the dimension of the coset $G/H$ does not work. 

In fact, this argument works in any dimension. In $D=1+n$ spacetime dimensions, the $SL(n,\mathbb{R})$ of the asymptotic symmetry group is broken down to $SO(n,\mathbb{R})$, leaving $n(n+1)/2-1$ broken shear transformations. Now out of those $n$ can be related by unbroken spatial translations, leaving $n(n-1)/2-1 = D(D-3)/2$ Goldstone modes, which is the number of degrees of freedom of a graviton in $D=1+n$ spacetime dimensions.

Following \cite{Ferreira:2016hee}, we can extend the transformation in Eq.(\ref{vgamma}) to higher order in $\gamma$. To next order in $\gamma$, we obtain
\beq\label{vgamma2}
\delta \gamma_{ij}= \p_i \xi_j +\p_j\xi_i - \frac{2}{3}\p\cdot\xi \delta_{ij} +\xi \cdot \p \gamma_{ij} +\mathcal{O}(\gamma^2)\, .
\eeq
In general, for a spontaneously broken symmetry generated by the charge, $Q$, there exists an order parameter $[Q,\mathcal{O}]=-i\delta \mathcal{O}$ for some operator $\mathcal{O}$ \cite{Brauner:2010wm}. Using now $\xi=\xi_i^L$ given by Eq.(\ref{xiL}) in Eq.(\ref{vgamma2}), and substituting the expression into the standard expression for the N\"{o}ther charge  
\beq
Q(\xi) = \frac{1}{2} \int d^3 x \Pi_\gamma^{ij}\delta \gamma_{ij}~ + ~ h.c.~,
\eeq
where $\Pi_\gamma$ is the canonical momentum, one finds that 
\beq
\langle 0| [Q,\gamma_{ij}]|0\rangle=i \gamma_{ij}^L\, ,
\eeq
by using the standard canonical commutation relations. This is as expected when $Q$ is the charge generating a spontaneously broken symmetry transformation and $\gamma_{ij}^L$ is the order parameter, and we conclude that $Q$ generates the shift $\delta\gamma_{ij} =-\gamma_{ij}^L$ in $\gamma_{ij}$ \cite{Ferreira:2016hee}.  

We saw above that rotational invariance is unbroken at zeroth order in $\gamma$. However, since the order parameter is not invariant under rotations, it breaks rotational symmetry. Rotational symmetry is, therefore, broken at linear order in $\gamma$. In fact, this can be seen from taking $\xi_i =-\omega_{ij}x^j$, with $\omega_{ij} = -\omega_{ji}$ corresponding to a rotation. In this case, we have, when the rotation acts on $\gamma_{ij}$ at higher order, $\delta \gamma_{ij} = \omega_{jl} \gamma_{li} + \omega_{il} \gamma_{lj}$, and so $\p_i \delta \gamma_{ij} = \p_i \omega _{il}\gamma_{lj}\neq 0$, such that $\delta\gamma_{ij}$ generated by a rotation acting on $\gamma_{ij}(x)$ is not transverse, meaning that the rotations leave the transverse and traceless gauge at linear order in $\gamma$. Thus, at second order, rotational symmetry is nonlinearly realized on the Goldstone mode. By using the fact that the angular momentum operator takes the form $J^i \propto \int d^3 x\epsilon^{ijk} \gamma_{mk}\dot\gamma_{ml}$, one also finds $[J^i,\gamma_{ij}] \neq 0$, as expected for the generator of a broken symmetry. So, while the anisotropic scaling symmetries, the shear transformations, are spontaneously broken at zeroth order in $\gamma$, their breaking induces a breaking of rotational symmetry at higher order.

We have found that the order parameter is a symmetric and traceless tensor, which breaks rotational invariance, like in the case of a Nematic liquid crystal, although the spontaneous symmetry-breaking pattern is different. In Nematic liquid crystals, the rotational symmetry is spontaneously broken, while anisotropic scaling is explicitly broken. 

The Goldstone modes are the long-range spatial fluctuations of the order parameter $\gamma_{ij}^L$ . Thus, the soft modes are Goldstones of the spontaneously broken asymptotic symmetries. This also means that the Lagrangian for the Goldstone modes is 
of the general form \cite{Naegels:2021ivf}
\beq
\mathcal{L}(\pi) \propto  g_{ab} \p_\mu \pi^a\p^{\mu}\pi^b +\mathcal{O}(\p^4),
\eeq
where the metric $g$ is the most general metric invariant under the isometries of the group $G$, which is broken to $G/H$. The most general tensor invariant under the isometries of $\mathbb{R}^n$ is the rotational invariant Kronecker-delta function, so in our case, the indices of the Goldstone mode are contracted with $\delta_{ij}$, while the space-time indices are contracted with the space-time de Sitter metric, such that the Lagrangian becomes to leading order
\beq
\mathcal{L}(\gamma) \propto -\dot\gamma_{ij}(x)\dot\gamma_{ij}(x)+a^{-2} \p_l\gamma_{ij}(x)\p_l\gamma_{ij}(x) ~,
\eeq
which happens to agree with the quadratic action of (soft) gravitons in de Sitter spacetimes.

\section{Symmetry breaking: local versus global observers}\label{globalvslocal}
The N{\"o}ther charge generating the spontaneously broken asymptotic symmetry transformation
\beq
Q(\xi) = \frac{1}{2} \int d^3 x \Pi_\gamma^{ij}\delta \gamma_{ij}~ + ~ h.c.~,
\eeq
where $\Pi_\gamma$ is the canonical momentum, was derived in \cite{Ferreira:2016hee}. Using Eq.(\ref{vgamma}) and Eq.(\ref{xiL}), it was shown that the charge generating the shift in the gravitational field is equivalent to adding a soft (long) mode, such that in the shifted vacuum, 
\beq
\left| 0' \right>= e^{iQ_{\gamma_L}} \left| 0 \right>  \equiv \left| \gamma_L \right>
\eeq
the expectation value of the graviton becomes
\ba
\left< 0' \right| \gamma_{ij} \left| 0' \right> &\equiv& \left< \gamma_L \right| \gamma_{ij} \left| \gamma_L \right> = \left< 0\right| e^{-iQ_{{\gamma}_L}} \gamma_{ij} e^{iQ_{{\gamma}_L}}\left| 0\right>\nonumber\\ &=&- i  \left< 0\right| \left[Q_{{\gamma}_L},\gamma_{ij}\right] \left| 0\right> = \gamma_{ij}^L
\ea
assuming that $\left< 0 \right| \gamma_{ij}\left| 0\right>=0$, and defining $Q(\xi_i^L)\equiv Q_{{\gamma}_L}$.  Thus, we see that the vacuum is infinitely degenerate, with each of the $\left| \gamma_L \right>$ states, related by symmetry transformations, representing different vacuum states of spontaneously broken anisotropic scaling symmetry with different values of $\gamma_{ij}^L$.

Following an argument similar to the standard one for $U(1)$ symmetry breaking in \cite{Wilczek:2012jt},  we can now take a superposition of all the $\left| \gamma_L \right>$ states, which have no preference for a specific $\gamma_L$ value by integrating it out
\beq
|\Psi_{\text{sym}}\rangle=\frac{1}{\mathcal{N}} \int \mathcal{D} \gamma_{i j}^{L} \psi\left[\gamma_{i j}^{L}\right]\left|\gamma_L\right\rangle ~.
\eeq
with $\psi\left[\gamma_{i j}^{L}\right]$ even in $\gamma_{i j}^{L} \rightarrow-\gamma_{i j}^{L}$, such that 
\beq
\left\langle\Psi_{\text {sym }}\right| \gamma_{i j}(x)\left|\Psi_{\text {sym }}\right\rangle=0
\eeq
in the symmetric state. The wave function
\beq
\psi\left[\gamma_{i j}^L\right] =  \left\langle\gamma_L \mid \Psi_{\text {sym }}\right\rangle
\eeq
can be interpreted as the soft part of the wave function of the universe. Here, I used that the $|\gamma^L\rangle$ are orthogonal in the large volume limit. To understand this and what I mean by large volume limit, consider dividing a region of the spacelike future infinity into $N$ patches that are uncorrelated, except along the long-range order parameter. One could think of this as $N$ Hubble patches. The wave function of the region then factorizes into a product of the wave functions in each patch
\beq
\Psi_{\gamma_L}\left(x_{1}, x_{2}, \ldots\right) \approx \prod_{i=1}^{N} \psi_{\gamma_{L}}\left(x_{i}\right)
\eeq
using $\left|\left\langle\psi_{\gamma_{L}}\left(x_{i}\right) \mid \psi_{\gamma_{L}^{\prime}}\left(x_{i}\right)\right\rangle\right|=\left|\epsilon_{\gamma_{L} {\gamma_{L}}^{\prime}}\right|<1$
and that the patches are highly uncorrelated except for along the order parameter, we have
\ba
\left\langle\Psi_{\gamma_{L}} \mid \Psi_{\gamma_{L}^{\prime}}\right\rangle
&\approx& \prod_{i=1}^{N}\left\langle\psi_{\gamma_{L}}\left(x_{i}\right) \mid \psi_{\gamma_{L}^{\prime}}\left(x_{i}\right)\right\rangle \nonumber\\ &\propto& \left(\epsilon_{\gamma_{L} {\gamma_{L}}^{\prime}}\right)^{N}\rightarrow 0\quad \text{for}\quad  N \rightarrow \infty\, .
\ea
Similarly, for any finite combination of $n$ local observables of the form $\mathcal{O}\left(x_{i}\right)$, we have assuming cluster decomposition
\ba
&&\left\langle\Psi_{\gamma_{L}^{\prime}}\right| \mathcal{O}_{1}\left(x_{a}\right) \mathcal{O}_{2}\left(x_{b}\right) \cdots \mathcal{O}_{n}\left(x_{j}\right)\left|\Psi_{\gamma_{L}}\right\rangle \\ &\propto& \left(\epsilon_{\gamma_{L} {\gamma_{L}}^\prime}\right)^{N-n}\prod_{i=1}^n\langle\mathcal{O}_i\rangle   \rightarrow 0 \quad \text{for} \quad\gamma_{L}^{\prime} \neq \gamma_{L}\, ,\quad N \rightarrow \infty\, .\nonumber
\ea
Thus, the local operators are approximately diagonal in the $\left| \Psi_{\gamma_{L}}\right>$ basis, which means that any local measurement, including the effect of the environment,  will project to one of the states $\left|\gamma_L \right>$, approximated by $\left| \Psi_{\gamma_{L}}\right>$, and so only one of the $\left|\gamma_L\right>$ states matters. Locally, however, the $\left|\gamma_L \right>$ states are not diagonal, and one can not distinguish between them, so the local vacuum state is well described by the $|\Psi_{\text{sym}}\rangle$ state. 

The point is that if one takes the ensemble average of 
$\left<\gamma_{ij}\right> $ in many local vacuum states, one will find that it vanishes. Therefore, the local state is well represented by the symmetric state, $\left|\Psi_{\text{sym}}\right>$, which is a superposition of all possible values of $\gamma_L$. In this local state, $\gamma_L$ has no physical meaning, as the state does not depend on it. On the other hand, when comparing $\left<\gamma_{ij}\right> $  in different local patches, one will see that they correlate in the infinite volume limit. The infinite volume is, therefore, well described by a single choice of $\gamma_{ij}^L(x)$ and the symmetry is broken by the global state, which is well represented by $\left|\Psi_{\gamma_{L}}\right>$.  It is the act of measurement, or equivalently, the environment, which breaks the symmetry. Thus, once we include observers in the discussion of de Sitter space, the symmetry is spontaneously broken. This emphasizes the old mantra that in any discussion of a quantum system, the observers are part of the system \cite{bohr}.

\section{Inflation and spontaneously broken dilatation}\label{inflation}
In the case of inflation, the broken time-translation symmetry of de Sitter space implies that there is a scalar curvature perturbation, $\zeta$, in addition to the usual two graviton modes. This mode corresponds to the trace part of the metric, and Eq.(\ref{metricvariation}), therefore generalizes to
\beq \label{metricvariation2}
\delta \left(e^{2\zeta}\left[e^{\gamma}\right]_{ij} \right)=\cL_\xi \left( e^{2\zeta}\left[e^{\gamma}\right]_{ij}\right)~.
\eeq
From this, we deduce that at the lowest order 
\beq
\delta \zeta = \frac{1}{3}\partial\cdot\xi \, .
\eeq
Thus, a dilatation $\xi = c x^i$ will induce a constant shift $\delta \zeta = c$ corresponding to a soft mode of the curvature perturbation. We therefore conclude that dilatations generate a soft mode of the curvature perturbation, with
\beq
\xi^i_L = \zeta_L x^i
\eeq
implying $\delta \zeta = \zeta_L$. Just like before, we can therefore identify the soft $\zeta$ modes as Goldstone modes of the spontaneously broken dilatation symmetry, and again write the Lagrangian on the general form
\beq
\mathcal{L}(\zeta) \propto -\dot\zeta^2 +a^{-2} (\partial \zeta)^2
\eeq
which agrees with correct quadratic action of $\zeta$ after some integrations by parts \cite{Maldacena:2002vr}. This is an elegant, different way of understanding why the action looks particularly simple in the comoving gauge. In inflation in four spacetime dimensions, we therefore have three Goldstone modes. The two from the spontaneously broken anisotropic scale transformation, of which the soft tensor modes, $\gamma_{ij}^L$ are the Goldstones, and one from the breaking of dilations, of which the curvature perturbation, $\zeta_L$, is the Goldstone mode.

\section{Examples}\label{examples}
In order to further establish our understanding of the soft graviton modes as the Goldstone modes of spontaneously broken asymptotic symmetries, and to gain some further physical intuition, let me connect this perspective to some known results from the literature.

\subsection{Three-point consistency relation}
In \cite{Ferreira:2016hee} it was shown that the explicit construction of the charge can be used to calculate the three-point function in the squeezed limit. Consider
\beq
\begin{aligned}
\left\langle 0^{\prime}\right| \gamma_{L} \gamma_{S} \gamma_{S}\left|0^{\prime}\right\rangle & =\langle 0| e^{-i Q} \gamma_{L} \gamma_{S} \gamma_{S} e^{i Q}|0\rangle \\
& =\langle 0| \gamma_{L} \gamma_{S} \gamma_{S}|0\rangle-i\langle 0|\left[Q, \gamma_{L} \gamma_{S} \gamma_{S}\right]|0\rangle+\ldots
\end{aligned}
\eeq
An explicit evaluation of the commutator reproduces the Maldacena consistency relation
\beq
\langle 0|\left[Q, \gamma_{L} \gamma_{S} \gamma_{S}\right]|0\rangle=\left\langle\gamma_{L} \gamma_{L}\right\rangle k^{2} \frac{\partial}{\partial k^{2}}\left\langle\gamma_{S} \gamma_{S}\right\rangle
\eeq
where $\gamma_L$ is the long wavelength soft mode and $\gamma_S$ is a short, UV, mode.

\subsection{Infrared loop corrections}
Similarly, one can also check that the IR one-loop corrections of the two-point correlation function of short scalar modes during inflation can be computed by
\beq
\left\langle 0^{\prime}\right| \zeta_{S} \zeta_{S}\left|0^{\prime}\right\rangle=\langle 0| e^{-i Q} \zeta_{S} \zeta_{S} e^{i Q}|0\rangle
\eeq
where an explicit evaluation and averaging over long (soft) modes reproduces the result of \cite{Giddings:2010nc} (see \cite{Ferreira:2016hee}). Similar relations hold for tensor modes as well.

\subsection{Orthogonalization of states}
We showed earlier that the $\left| \gamma_L \right>$ states become orthogonal in the large volume limit. Since de Sitter spacetimes are expanding, this is equivalent to the late time limit. We can therefore also use the results of \cite{Ferreira:2016hee} to understand the timescale of orthogonalization of the $\left| \gamma_L \right>$ states. In \cite{Ferreira:2016hee} it was shown that
\beq
\langle 0^{\prime} \mid 0\rangle=\langle 0| e^{-i Q}|0\rangle = 1- \frac{1}{2}\frac{8 \pi^2}{45} \langle \gamma_L \gamma_L\rangle +\dots
\eeq
Using that the variance of the soft modes in de Sitter spacetimes grows as 
\beq
\langle \gamma_L \gamma_L\rangle \simeq \frac{1}{2\pi^2}\frac{H}{M_p}\log(a(t))
\eeq
where $a(t)$ is the scale factor, we have that the states orthogonalize on the time scale at which 
\beq
\frac{H}{M_p}\log(a(t)) \sim \mathcal{O}(1)~.
\eeq
When using the relation $\log(a(t)) = Ht$ for de Sitter space, one finds that this is the equivalent of the Page time for de Sitter space, $t\sim 
M_p^2/H^3\sim R_{dS} S_{dS} $, where  $R_{dS}$ is the de Sitter radius and $S_{dS}$ is the de Sitter entropy \cite{Ferreira:2016hee,Giddings:2011ze,Giddings:2010nc}. A related argument was made in the context of black holes and the information paradox already in \cite{Giddings:2007ie}. 

Thus, in de Sitter spacetimes, the Page time can be understood as the time scale on which we start to probe the broken phase, and can no longer characterize our state by the symmetric state.

\section{Conclusions and outlook}\label{conclusions}
I have shown that the soft graviton mode in de Sitter spacetimes can be viewed as the Goldstone mode of the spontaneously broken asymptotic symmetry group of de Sitter space. I showed that the argument holds in any dimension and generalizes straightforwardly to inflationary, quasi–de Sitter spacetimes. 

Through examples, I discussed how this is just another formulation of the usual consistency relations and can be viewed as the de Sitter version of the equivalence between soft theorems and asymptotic symmetries, which, when also thought of as a gravitational memory effect \cite{Ferreira:2017ogo}, completes the triangle relating asymptotic symmetries, soft theorems and gravitational memory \cite{Strominger:2017zoo}. 

There are different applications and extensions of my work that one could consider. Extending the Lagrangian of the Goldstone mode to higher orders, one has a different approach to an effective field theory of inflation or dark energy. In this context, it could also be interesting to extend our formalism to modified gravity theories or massive gravity theories. In addition, it is known that the charge, $Q$, has a dual Conformal Field Theory (CFT) interpretation as a topological charge \cite{Kehagias:2017rpe}, and it would be interesting to explore the CFT dual of the spontaneous symmetry breaking. Finally, it would be interesting if a {\it de Sitter liquid crystal}, with the same symmetry-breaking pattern as discussed here, could be realized in the laboratory, as an analogue system of de Sitter spacetimes and eternal inflation.

\subsection*{Acknowledgments}

I would like to thank Alex Kehagias for the stimulating discussions and useful comments on the draft, which helped me finish this work.

\end{document}